\documentclass[prd,11pt,tightenlines,nofootinbib,superscriptaddress]{revtex4}
\usepackage{amsfonts,amssymb,amsthm,bbm}

\usepackage{amsmath,amssymb,bbm}
\usepackage{amsfonts}
\newcommand{\N}{{\mathbb N}}

\newcommand{\cE}{{\mathcal E}}
\newcommand{\cF}{{\mathcal F}}
\newcommand{\cG}{{\mathcal G}}
\newcommand{\cJ}{{\mathcal J}}

\newcommand{\cH}{{\mathcal H}}

\newcommand{\cN}{{\mathcal N}}

\newcommand{\cD}{{\mathcal D}}
\newcommand{\cC}{{\mathcal C}}
\newcommand{\cS}{{\mathcal S}}
\newcommand{\SU}{\mathrm{SU}}
\newcommand{\U}{\mathrm{U}}

\newcommand{\vJ}{\vec{J}}
\newcommand{\w}{\wedge}

\newcommand{\be}{\begin{equation}}
\newcommand{\ee}{\end{equation}}
\newcommand{\beq}{\begin{eqnarray}}
\newcommand{\eeq}{\end{eqnarray}}
\newcommand{\bea}{\begin{eqnarray}}
\newcommand{\eea}{\end{eqnarray}}
\newcommand{\nn}{\nonumber}

\newcommand{\bin} [2] {\left (\begin{array}{c}#2\\#1\end{array} \right ) }

\newcommand{\su}{{\mathfrak su}}
\newcommand{\osp}{{\mathfrak osp}}
\renewcommand{\u}{{\mathfrak u}}

\newcommand{\la}{\langle}
\newcommand{\ra}{\rangle}

\newcommand{\f}{\frac}
\newcommand{\vphi}{\varphi}

\newcommand{\Ref}[1]{(\ref{#1})}

\def\nn{\nonumber}
\def\pp{\partial}

\def\arr{\rightarrow}
\def\tF{\widetilde{F}}
\def\tG{\widetilde{G}}

\def\cF{{\cal F}}

\begin{document}

\title{The Fine Structure of $\SU(2)$ Intertwiners from $\U(N)$ Representations}

\author{Laurent Freidel}
\affiliation{ Perimeter Institute for Theoretical Physics,
Waterloo, N2L-2Y5, Ontario, Canada.}
\author{Etera R. Livine}
\affiliation{Laboratoire de Physique, ENS Lyon, CNRS-UMR 5672, 46 All\'ee d'Italie, Lyon 69007, France.}

\date{November 18, 2009}

\begin{abstract}
In this work we study the Hilbert space space of $N$-valent $\SU(2)$ intertwiners with fixed total spin,
which can be identified, at the classical level, with a space of convex polyhedra with $N$ face and
fixed total boundary area.
We show that this Hilbert space  provides, quite remarkably,  an irreducible representation of the $\U(N)$ group.
This gives us therefore a precise identification of $\U(N)$ as a group of area preserving diffeomorphism of polyhedral spheres.
We use this results to get new closed formulae for the black hole entropy in loop quantum gravity.

\end{abstract}

\maketitle


\section{Introduction}

Loop quantum gravity is an attempt to quantize canonically and non-perturbatively general relativity. At the kinematical level, it defines the quantum states of 3d space geometry as spin networks. These spin network wavefunctions are cylindrical functionals of the Ashtekar-Barbero connection and diagonalize geometrical operators such as areas and volumes. A spin network is defined on a graph, with $\SU(2)$ representations (spins) attached to each edge and $\SU(2)$ invariant tensors (intertwiners) attached to each vertex, and can be interpreted as a quantized triangulated 3d geometry. Geometrically, the edges are dual to surfaces and the spins give their area, while vertices are dual to elementary chunks of 3d space and the intertwiners are supposed to describe their shape and volume.
At the dynamical level, one seeks to impose the Hamiltonian constraint implementing the invariance under space-time diffeomorphisms on the Hilbert space of spin network states. This dynamics encodes the evolution of the spin network state, that is of the spins and intertwiners and of the underlying graph itself.

In the present paper, we investigate the structure of the Hilbert space of intertwiners. As we said previously, these are the basic building blocks of spin network states and gluing intertwiners together generates the quantum 3d space. Understanding better the structure of the intertwiner space and its geometrical interpretation is necessary for a better understanding of loop gravity at both kinematical and dynamical level.

More specifically, we consider the space of $N$-valent intertwiners, that is of invariant tensor of $N$ spins ($\SU(2)$ representations). Such intertwiners can be thought dually as a region of 3d space with a (topologically) spherical boundary \cite{Barbieri,Coh2, Holomorph} . This boundary is punctured by the $N$ legs of the intertwiners: the boundary surface is made of $N$ elementary patches whose area is determined by  the spins carried by the intertwiner legs.

Here, we follow the approach initiated in  \cite{OH}, where it was shown that  the Hilbert
 space of $N$-valent intertwiners at fixed total boundary area  admits an SU$(N)$ action.
 Our main result consists in showing that this space is in fact an irreducible  representation (irrep) of $\U(N)$. We fully determine this $\U(N)$-irrep through its highest weight and give explicitly their dimensions and characters. We further interpret this $\U(N)$-action as the area-preserving diffeomorphisms acting on the punctured boundary surface at the quantum level. This is the main result of the paper. It goes in the same direction as interpreting $\SU(\infty)$ as the group of area-preserving diffeomorphisms of a surface in the semi-classical limit (see e.g. \cite{hoppe}).

Moreover, we relate the calculation of the dimensions of the intertwiner spaces through this new $\U(N)$ technique to the computation of black hole entropy in loop quantum gravity. In particular, we compare it to the other standard results in this field.
Then, we explain how to glue these intertwiner spaces together to describe the full Hilbert space of spin network states. This actually provides a dual point of view on spin networks with a shift of the degrees of freedom from the edges to the vertices of the graph.
Finally, we discuss the possible generalizations of our framework to gauge groups more complicated than $\SU(2)$, to supersymmetric gauge groups and to quantum groups.

\section{The $\U(N)$ Action on the Intertwiner Space}

In the following, we call $V^j$ the Hilbert space corresponding to the irreducible representation of $\SU(2)$ with spin $j\in\N/2$. Its dimension is $d_j=(2j+1)\in\N$.
Considering $N$ irreps of spins $j_1,..,j_N\,\in\N/2$, the corresponding space of intertwiners with $N$ legs is made of vectors in the tensor product of these representations which are invariant under the global $\SU(2)$ action:
\be
\cH_{j_1,..,j_N}\,\equiv\, \textrm{Inv}[V^{j_1}\otimes..\otimes V^{j_N}].
\ee
We are interested in describing the structure of the full space of $N$-leg intertwiners:
\be
\cH_N\,\equiv\,\bigoplus_{\{j_i\}}\cH_{j_1,..,j_N}.
\ee

\subsection{The Hidden $\U(N)$ structure of the Intertwiner Space}

The usual invariant operators that we consider on the intertwiner space to characterize the invariant states are the scalar product operators. Writing $\vJ^{(i)}$ for the three generators in the Lie algebra $\su(2)$, these Hermitian operators are simply the $\vJ^{(i)}\cdot \vJ^{(j)}$ for all couples of indices $(i,j)$. Interpreting the generators $\vJ^{(i)}$ as defining a quantum 3-vector associated to the $i$th patch on the boundary surface, the operators $\vJ^{(i)}\cdot \vJ^{(j)}$ will effectively measure the scalar product between these vectors. Of course the actual expectation value of $\vJ^{(i)}\cdot \vJ^{(j)}$ on an intertwiner state will usually deviate from the scalar product of the expectation values of the operators $\vJ^{(i)}$ and $\vJ^{(j)}$.

The standard issue with these scalar product operators is that they do not form a closed Lie algebra:
\be
[\vJ^{(i)}\cdot \vJ^{(j)},\vJ^{(i)}\cdot \vJ^{(k)}]=\,i\,\vJ^{(i)}\w(\vJ^{(j)}\cdot \vJ^{(k)}).
\ee
indeed the commutator of two scalar product operators gives a new operator cubic in the $\vJ$'s. This cubic operator can be interpreted as measuring the 3-volume generated by the three vectors $\vJ^{(i)}$, $\vJ^{(j)}$ and $\vJ^{(k)})$. Computing further commutators of these operators will generate higher and higher order operators in the $\vJ$'s.
This a priori infinite dimensional algebra has not yet been really studied (to our knowledge) and does not help us characterize the Hilbert space of intertwiners. Moreover, since the commutators of the scalar product operators do not close, we can not build coherent semi-classical states using this choice of invariant operators.

The alternative presented in \cite{OH} is to use the Schwinger representation of the $\su(2)$ algebra in term of a couple of harmonic oscillators:
$$
[a,a^\dag]=[b,b^\dag]=1,\qquad [a,b]=0,
$$
and we define the $\su(2)$ generators as quadratic operators in the $a,b$'s:
\be
J_z=\f12(a^\dag a-b^\dag b),\quad
J_+=a^\dag b, \quad
J_-=ab^\dag.
\ee
We also define the (half) total energy $E$:
\be
\cE=\f12(a^\dag a+b^\dag b)
\ee
It is straightforward to check that the commutation relations reproduce the expected $\su(2)$ structure:
$$
[J_z,J_\pm]=\pm J_\pm, \quad [J_+,J_-]=2J_z, \quad
[\cE,\vJ]=0.
$$
From these definitions, it is direct to identify the correspondence between the usual basis of the Hilbert space for harmonic oscillators and the standard basis of $\SU(2)$ irreps in term of the spin $j$ and magnetic momentum $m$~:
$$
|j,m\ra\,=|n_a,n_b\ra_{OH},
\qquad\textrm{with}\quad
j=\f12(n_a+n_b),\quad m=\f12(n_a-n_b).
$$
The total energy $\cE$ actually gives the spin $j$ and we check that the $\su(2)$ Casimir operator can indeed be simply expressed in term of $\cE$:
$$
\cC=\vec{J}^2=\cE(\cE+1).
$$

Considering intertwiners with $N$ legs, we need to take $N$ irreps of $\SU(2)$, so we use $2N$ oscillators $a_i,b_i$. Following  \cite{OH}, we define the quadratic operators acting on couples of punctures $(i,j)$:
\be
E_{ij}\equiv (a_i^\dag a_j+b_i^\dag b_j), \quad
E_{ij}^\dag=E_{ji}.
\ee
These operators commute with the global $\SU(2)$ transformations, so they legitimately define operators acting the intertwiner space:
\be
\forall i,j,\quad \left[\sum_k \vJ^{(k)}\,,\,E_{ij}\right]=0.
\ee
Moreover it is easy to check that the new operators form a closed $\u(N)$ Lie algebra:
\be
[E_{ij},E_{kl}]\,=\,
\delta_{jk}E_{il}-\delta_{il}E_{kj}.
\ee
The diagonal operators $E_i\equiv E_{ii}$ form the (abelian) Cartan sub-algebra. Their value on a state gives twice the spin on the $i$th leg, $2j_i$. The off-diagonal operators $E_{ij}$ define the lowering and raising operators in $\u(N)$.

The main difference between the $E_{ij}$ operators and the scalar product operators $\vJ^{(i)}\cdot \vJ^{(j)}$ is that the $E$'s are quadratic in the $a,b$'s while the $\vJ\cdot\vJ$ are quartic. Resulting the commutator of two $E$ operators is once again quadratic in the $a,b$'s and their commutators close while the commutator of two $\vJ\cdot\vJ$ operators becomes of order 6 in the $a,b$'s. Somehow, we have managed to take the square-root of the scalar product operators. More precisely, as was shown in \cite{OH}, it is possible to express the operator $\vJ^{(i)}\cdot \vJ^{(j)}$ as a quadratic polynomial of $E_{ij}, E_{ji},E_i,E_j$~:
\be
\vJ^{(i)}\cdot \vJ^{(j)}=\f12E_{ij}^\dag E_{ij} -\f14E_iE_j-\f12E_i
\ee

Finally, the $\u(1)$ Casimir operator, which generates the global $\U(1)$ phase in $\U(N)$, is $E\equiv\sum_i E_i$. Its value on a state gives twice the sum of the spins on all legs, $2\sum_i j_i$. We interpret $E$ as measuring (twice) the total area of the boundary surface around the intertwiner. Indeed $E$ commutes with all the other $\u(N)$ operators, $[E,E_{ij}]=0$, and we can interpret the newly defined $\U(N)$ as the {\it group of area-preserving diffeomorphisms} of boundary surface considered as a discretized sphere. Indeed $\U(N)$ transformations will act on the space of intertwiners and will deform the intertwiners but keeping the total area $E$ fixed. Hence, the natural definition of the area is our framework is:
\be
\textrm{Area}\,\equiv\,\sum_i j_i,
\ee
which is different from the standard loop quantum gravity area $\sum_i \sqrt{j_i(j_i+1)}$.

The next question is to describe the action of these $\U(N)$ transformations on the intertwiner states, that is to determine which $\U(N)$ representation we obtain. We address this issue below.

\subsection{Generating $\SU(N)$ from Harmonic Oscillators}
\label{proofweight}

Actually, constructing the $\u(N)$ Lie algebra from harmonic oscillators is a pretty standard mathematical construction.
Starting from $P$ sets of $N$ harmonic oscillators, $a^{(p)}_i$ with the labels $i$ running from 1 to $N$ and $p$ from 1 to $P$, commuting with each other, we can build a representation of the unitary group $\U(N)$. Indeed, defining the quadratic operators:
\be
E_{ij}=\sum_{p} a^{(p)\,\dag}_i a^{(p)}_j,
\ee
it is straightforward to check that their commutators form a $\u(N)$ Lie algebra:
\be
[E_{ij},E_{kl}]\,=\,
\delta_{jk}E_{il}-\delta_{il}E_{kj}.
\ee
The case relevant for the study of $\SU(2)$ intertwiners is given by $P=2$. The special case $P=N$ is the one usually considered in order to study the representations and the recoupling theory of $\U(N)$.
Here, the natural question which we would like to address is to identify the representations of $\u(N)$ that this construction induces.

Let us start by looking at the $P=1$ case, we can drop the $p$ index and we construct the $\u(N)$ Lie algebra from one set of $N$ oscillators:
\be
E_{ij}=a^\dag_i a_j,\qquad [a_k,a^\dag_l]=\delta_{kl}.
\ee
Using as previously the conventions $E_i\equiv E_{ii}$ and $E\equiv\sum_i^N E_i$ for the generators of the Cartan sub-algebra (diagonal elements), it is possible to check the generators constructed above satisfy the following constraint~\footnotemark:
\be
\label{many}
\forall i,\,\sum_j E_{ij}E_{ji}
=E_{i}(E+N-1).
\ee
\footnotetext{
Actually, this equation can be generalized to arbitrary $i,k$: $\quad\sum_j E_{ij}E_{jk}=(E+N-1)E_{ik}$.
}
If we further sum over the subscript $i$, we can an equation between Casimir operators:
\be
\label{one}
\sum_{i,j} E_{ij}E_{ji}=E(E+N-1).
\ee
$E$ is the generator of the $\U(1)$ phase of $\U(N)$,
while $\cC_2\,\equiv\,\sum_{i,j} E_{ij}E_{ji}$ is the quadratic Casimir operator (or quadratic Gel'fand invariant) of $\U(N)$.

To identify which representations are allowed, we apply this equation to the highest weight vector of the considered irreducible representation, i.e. $v$  such that $E_i\,v=l_i\,v$ with the weights $l_i\in\N$ (and $l_1\ge l_2\ge .. \ge l_N\ge0$) and $E_{ij}\,v=0$ for all $i<j$. Physically, the $l_i$ measure the energy of each oscillator. This allows to express the values of the $\U(N)$ Casimir operator in term of the eigenvalues $l_i$~:
$$
\forall i,\quad
\sum_j E_{ij}E_{ji}\, v= (E_i)^2\,v + \sum_{i<j}[E_{ij},E_{ji}]\,v
\,=\,\left[l_i^2 + \sum_{j>i}(l_i-l_j)\right]\, v.
$$
Evaluating the equation \Ref{many} provides a set of $N$ (quadratic) constraints on the weights:
\be
\forall i,\quad
\left[l_i^2 +l_i(N-i)- \sum_{j>i}l_j\right]= l_i(L+N-1),
\quad\textrm{or equivalently}\quad
l_i^2 -l_i(L+i-1)- \sum_{j>i}l_j= 0,
\ee
where $L=\sum_i l_i$ is the eigenvalue of the $\U(1)$ Casimir $E$. For $i=1$, this simplifies to $(l_1-L)(l_1+1)=0$. Taking into account that $l_i\ge 0$ for all $i$'s, we necessarily have $l_1=L$, which in turn implies, since $L=\sum_i l_i$,  that all $l_i=0$ for $i\ge 2$. We can check that this provides a solution to all constraints.

Instead of dealing with these $N$ constraints, we could have proceeded directly from the Casimir equation \Ref{one}. Indeed, we can coimpute the quadratic Casimir on the highest weight:
\be
\cC_2\,=\, \sum_i \left[l_i^2 + \sum_{j>i}(l_i-l_j)\right]
\,=\,\sum_i l_i(l_i+N+1-2i).
\ee
Equating this with the value of $E(E+N-1)$ gives:
$$
\sum_i l_i(L-l_i+2(i-1))\,=0.
$$
Since $0\le l_i \le L$ for all $i$'s, all the terms of this sum are positive and thus must all vanish. This implies that for each $i$, we have $l_i=0$ or $l_i=L+2(i-1)$. But since $l_{i}$  is always smaller than $L$ by definition, the only solution is $l_1=L$ and $l_j=0$ for $j\ge 2$.

Finally the only representations of $\u(N)$ induced by using a single set of $N$ harmonic oscillators  are the ones with highest weight $[L,0,0,..,0]$.
These highest weight representations are the symmetric representations of $\U(N)$, i.e. the ones with Young tableaux made of a single row with $L$ boxes.

We now show, considering a representation of $\U(N)$ constructed from $P$ sets of $N$ oscillators, that we obtain a tensor product of $P$ such symmetric representations, that is a Young tableaux made of $P$ rows at most. For $P=2$, this leads to highest weights $[l_1,l_2,0,0,..,0]$, or equivalently to Young tableaux with one or two rows. And so on, until we use the full $N$ sets of $N$ oscillators and obtain all possible irreducible representations of $\U(N)$.

\subsection{Identifying the Highest Weight}

We now come back to our case studying $\SU(2)$-invariant operators in term of the harmonic oscillators. This corresponds to the case $P=2$, but we need to take into account the further requirement of $\SU(2)$-invariance. In order to translate this requirement in a constraint on the $\U(N)$ representations, we write the new equations on the Casimir operators and apply them to the highest weight vector $v$ with eigenvalues $l_i$, $E_i\,v=l_i\,v$. It is straightforward to check the new set of constraints:
\be
\forall i,\quad
\sum_j E_{ij}E_{ji}=
E_i(\f{E}{2} +N-2)
\,+\,
2\left[
J^{(i)}_zJ_z +\f12 J^{(i)}_+J_- +\f12 J^{(i)}_-J_+
\right],
\ee
or if we also sum over the subscript $i$~:
\be
\sum_{i,j} E_{ij}E_{ji}=
E(\f{E}{2} +N-2)
\,+\,
2 \vec{J}\cdot\vec{J}
\ee
where $\vec{J} =\sum_{i}J^{(i)}$.
This latter equation relates the Casimirs operators of $\SU(N)$, $\U(1)$ and $\SU(2)$.

We focus on the representation of $\SU(N)$ carrying intertwiners, i.e. which are $\SU(2)$-invariant with $\vec{J}=0$. In this case, the highest weight vector satisfies the following constraints:
\be
\forall i,\,
\left[l_i^2 + \sum_{j>i}(l_i-l_j)\right]=
l_i\left(\f{L}{2} +N-2\right),
\,\textrm{or equivalently}\quad
l_i^2-\left(\f{L}2+i-2\right)l_i-\sum_{j>i}l_j=0.
\ee
Computing this constraint for $i=1$ gives $(l_1+2)(l_1-L/2)=0$, and thus necessarily $l_1=L/2$ since the weights $l_i$ are always positive. Then applying the formula to the case $i=2$ leads to
$(l_2+1)(l_2-L/2)=0$,  and thus necessarily $l_2=L/2$. Then all the others coefficients vanish, $l_3=l_4=..=l_N=0$.
Finally, we are left with the representations of highest weight $[l,l,0,..,0]$, with two rows of equal length in the Young tableau.
We see the we get a representation that has two  non-trivial weight  as promised in the prevous section.
The requirement that  these two weight are equal, $l_1=l_2$ comes from the $\SU(2)$-invariance.
Looking at this highest weight vector from the point of view of $\SU(2)$ intertwiners, it has a very simple interpretation, since the value $l_i$ is actually  the value of the spin $2j_i$ labeling the $\su(2)$ representation attached to the $i$-th leg of the intertwiner. Thus the highest weight vector $[l,l,0,..,0]$ corresponds to a bivalent intertwiner with $j_1=j_2=l/2$.

Actually, we can generalize this calculation to non-trivial values of the $\SU(2)$ Casimir operator. These are characterized in term of the overall spin $\cJ\in\N/2$, $\vec{J}\cdot\vec{J}=\cJ(\cJ+1)$. Writing the Casimir equation for a highest weight vector $[l_1,l_2,0,..,0]$, we get:
\be
(l_1-l_2)\left(\f{l_1-l_2}{2}+1\right)=2\cJ(\cJ+1), \qquad\textrm{thus}\quad
\cJ=\f{l_1-l_2}{2}.
\ee
This actually corresponds to the smallest $\SU(2)$ representation in the tensor product of the two $\SU(2)$ representations with spins $j_i=l_i/2$.

\medskip

At the end of the day, we have shown that the space of  $N$-leg intertwiners  for a fixed total area $l$
carries an irreducible  representation of $\U(N)$.
What we will show in the next section by computing the respective dimension is that   the space of  $N$-leg intertwiners having a fixed total area $l$ is in fact {\it isomorphic} an irreducible  representation of $\U(N)$:
\be\label{iso}
R^l_N\,=\,\bigoplus_{\sum_ij_i\,=l}\cH_{j_1,..,j_N}.
\ee
$R^l_N$ carries an irreducible representation of $\U(N)$ with highest weight $[l,l,0,0,..]$ with two equal non-trivial eigenvalues for the Cartan generators $E_i$. This highest weight vector describes a bivalent intertwiner between two copies of the same $\SU(2)$ representation of spin $j=l/2$. This bivalent intertwiner can be interpreted as a completely squeezed sphere, made with only two pacthes. Then we can act with $\u(N)$ operators to reach all the other $N$-leg intertwiners with the same total area.

We end up this section with two remarks. First, we underline the fact that the $\SU(2)$ representation on the legs of the intertwiners can be trivial, with $j_i=0$. This will be specially relevant when discussing (black hole) entropy. Second, using the hook formula for Young tableaux\footnotemark, we can compute the dimension of the $\U(N)$ representation $R^l_N$~:
\be
\dim_N[l]\,\equiv\,\dim \,R^l_N
\,=\,\f{1}{l+1}\bin{l}{N+l-1}\bin{l}{N+l-2},
\ee
in term of binomial coefficients.
This actually gives the total number of intertwiners with $N$ legs and for a fixed total area $l=\sum_i j_i$ including the possibility of trivial $\SU(2)$ irreps.
\footnotetext{
For a more general highest  $[l_1,l_2,0,..,0]$, corresponding to a non-trivial overall $\SU(2)$ spin $\cJ=(l_1-l_2)/2$, the dimension of the $\U(N)$ representation becomes:
$$
\dim_N[l_1,l_2]
\,=\,
\f{l_1-l_2+1}{l_1+1}\bin{l_1}{N+l_1-1}\bin{l_2}{N+l_2-2}.
$$
}
%

\subsection{The Generating Functional(s) for the Number of Intertwiners}
So far what we have shown in the next section is the fact that $R^l_N$ is an invariant subset of the intertwinner space $\bigoplus_{\sum_ij_i\,=l} \cH_{j_1,..,j_N}$.
In order to show the equality we ned to check that
the dimension of the intertwiner space is correctly given by $\dim_N[l]$.

More precisely, we define the dimensions of the intertwiners for fixed $\SU(2)$ irreps on their legs. These dimensions can be expressed in term of group integrals:
\be
\dim_0[j_1,..,j_N]
\,\equiv\,
\dim\,\cH_{j_1,..,j_N}
\,=\,
\int dg \prod_i^N\chi_{j_i}(g),
\ee
where $dg$ is the normalized Haar measure and $\chi_j$ the character in the irrep of spin $j$. Next we define the total number of intertwiners between $N$ punctures for a fixed sum of spins $l=\sum_i^N j_i$:
$$
\cN^{(N)}[l]\,\equiv\,\sum_{j_1+..+j_N=l}  \dim_0[j_1,..,j_N],
$$
and our goal is to check that $\cN^{(N)}[l]=\dim_N[l]$ as expected. This relation actually extend beyond the simple calculation of the dimension of the intertwiner space and can be generalized to $\U(N)$ characters. We give the details in appendix.

To this purpose, we compare their respective generating functionals at fixed number of legs $N$ but summing over the total area $l$:
\be
F_N(t)=\sum_{l\in\N} t^{2l}\dim_N[l],\qquad
\tF_N(t)=\sum_{l\in\N} t^{2l}\cN^{(N)}[l].
\ee
We point out that the total area $l$ is necessarily an integer due to the parity condition between spins for the existence of an intertwiner. On the one hand, we write a simple recursion relation for for $\dim_N[l]$:
\be
(l+1)(l+2)\,\dim_N[l+1]\,=\,(N+l)(N+l-1)\,\dim_N[l].
\ee
This leads to a differential equation on the corresponding generating functional:
\be
\Delta^{(N)}_t\,F_N(t)=0
\qquad\textrm{with}\quad
\Delta^{(N)}_t\,\equiv\, \f14(1-t^2)(t\pp_t^2 +3\pp_t)\,-N(N-1)t -(N-1)t^2\pp_t.
\ee
This second order differential equation implies that $F_N$ is a hypergeometric function up to a change of variable in $t$.

On the other hand, we can express $\tF_N$ as an integral using the explicit definition of the $\SU(2)$ characters~\footnotemark:
\beq
\label{FN}
\tF_N(t)\equiv \sum_{J\in\N/2}t^{2J} I^{(N)}[J]
&=&
\int dg\, \left[
\sum_j t^{2j}\chi_j(g)
\right]^N
\,=\,
\f2\pi\int_0^\pi d\theta\,\f{\sin^2\theta}{(1-2t\cos\theta+t^2)^N}, \\
&=&
\f2\pi\int_0^\pi d\theta\,\f{\sin^2\theta}{(t-e^{i\theta})^N(t-e^{-i\theta})^N}
\,=\,
\f2\pi\int_{-1}^{+1} dx\,\f{\sqrt{1-x^2}}{(1-2tx+t^2)^N}. \nn
\eeq
A straightforward calculation yields:
$$
\Delta^{(N)}_t \,\tF_N
\,=\,\f2\pi\int_0^\pi d\theta\,\Delta^{(N)}_t\,\f{\sin^2\theta}{(1-2t\cos\theta+t^2)^N}
\,=\,\f{N(1-t^2)}{\pi}\int_0^\pi d\theta\,
\pp_\theta\f{\sin^3\theta}{(1-2t\cos\theta+t^2)^{N+1}}=0,
$$
showing that $F_N$ and $\tF_N$ satisfy the same differential equation. Then checking by hand the initial conditions, $\dim_N[l=0]=1=\cN^{(N)}[l=0]$ and $\dim_N[1]=N(N-1)/2=\cN^{(N)}[1]$, we conclude that $F_N(t)=\tF_N(t)$.
This results implies that $dim(R^{l}_{N})$ is equal to the dimension of the space of intertwinner with fixed area
and this  proves the isomorphism (\ref{iso}) which is our main result.

\footnotetext{
We can compute explicitly these generating functionals for some values of the number of punctures $N$.  For instance, the case $N=2$ corresponts to purely bivalent intertwiners and we can easily evaluate the integral \Ref{FN}~:
$$
F_2(t)=\f{1}{1-t^2}=\sum_{l\in\N}t^{2l}
\qquad\textrm{which matches}\quad
\dim_2[l]=1,\,\forall l\in\N.
$$
In the case of $N=4$ punctures, a straightforward calculation gives:
$$
F_4(t)=\f{1+t^2}{(1-t^2)^5}=\sum_{l\in\N} \f{(l+1)(l+2)^2(l+3)}{12}t^{2l},
$$
which once again fits perfectly the formula given for the dimension of the $\U(4)$ representations $\dim_4[l]$. For $N=6$ punctures, we get:
$$
F_6(t)=\f{1+6t^2+6t^4+t^6}{(1-t^2)^9}
\,=\,\f{(1+t^2)(t^4+5t^2+1)}{(1-t^2)^9}.
$$
In general, it is possible to express this integral in term of the (associated) Legendre polynomials:
\be
\forall N\ge 4,\quad
F_N(t)=\f{1}{(N-1)(N-2)}\f{P^1_{N-2}\left(\f{1+t^2}{1-t^2}\right)}{t(1-t^2)^{N-1}},
\ee
where the $P^1_N$ polynomials  are the first derivatives of the Legendre polynomials up to factors (and can also be easily written as hypergeometric functions).
}

Now, we introduce the full generating functional by also performing the sum over the number $N$ of legs:
\be
\cF(u,t)=\sum_{N\in\N}u^N F_N(t)=\sum_{N,l} u^N t^{2l}\dim_N[l].
\ee
We can give it an integral expression:
\be
\label{cF}
\cF(u,t)\,=\,\int dg \,\sum_N \left[
u\,\sum_j t^{2j}\chi_j(g)
\right]^N
\,=\,
\f2\pi\int_0^\pi d\theta\,
\f{1-2t\cos\theta+t^2}{1-2t\cos\theta+t^2-u}.
\ee
This is a trigonometric integral which can be exactly computed for $u,t\sim 0$:
\be
\cF(u,t)=\f1{2t^2}\left[
t^2(u+2)-u^2+u-
u\,\f{t^2(t^2-2u-2)+(u-1)^2}{\sqrt{((1+t)^2-u)((1-t)^2-u)}}.
\right].
\ee
The value $u=1$ corresponds to the calculation of $\sum_N \dim_N[l]$. However, as we can see from the previous formula, the leading order behavior is divergent and imaginary with $\cF(1,t)\sim -i/t$ for $t\sim 0$. This reflects the simple fact that the sum $\sum_N \dim_N[l]$ is divergent (for all values of $l\in\N$).

Nevertheless, as soon as we set $0<u<1$, the function $\cF(u,t)$ is well-defined around $t\sim 0$ and we can expand it in power series in $t$. For instance, in the case $u=1/2$, we (or Maple) can compute its expansion:
\be
\cF(\f12,t)\,\underset{t\arr 0}{\sim}\,
2+2t^2+12t^4+88t^6+720t^8+6304t^{10}+\dots
\ee
One recognizes the sums $\sum_N \f1{2^N}\,\dim_N[J]$ for different values of $J$, which we can compute directly from the formula of the dimension of the $\U(N)$ dimension.

More generally, from the explicit expression of $\cF(u,t)$, for a fixed value of $0<u<1$, we can see by focusing on the square-root factor that the first pole in $t$ is given by $t_c=1-\sqrt{u}$. This gives the leading order asymptotics for large total area $l$ of the sums over $N$:
\be
\log\,\sum_N u^N \dim_N[l]\,\underset{l\arr\infty}{\sim}\,
-2l \log(1-\sqrt{u}).
\ee
For $u=\f12$, we get a leading order behavior in $l\log(6+4\sqrt{2})$. Moreover, this formula confirms that the series diverge when $u\arr 1$.

In the next section, we apply these calculations to black hole entropy in loop quantum gravity.

\section{About Intertwiner Counting and Black Hole Entropy}

\subsection{What should we count?}

We have computed the dimension of the intertwiner space $\dim_N[l]=\sum_{\sum_i j_i=l}\dim_0[j_1,..,j_N]$ for $N$ legs and a fixed total area $l$. This calculation seems to be related to black hole entropy in loop quantum gravity. Nevertheless, we point out several discrepancies:

\begin{itemize}

\item The standard framework for isolated horizon in loop quantum gravity \cite{lqgBH} counts the states of a boundary $\U(1)$ Chern-Simons theory, that is the number of $\U(1)$ intertwiners compatible with the bulk data (the total area). However, recent work has generalized these arguments to a boundary $\SU(2)$  Chern-Simons theory and counting of $\SU(2)$ intertwiners \cite{alej,spain}. This point of view had been been emphasized in several other works (see e.g. \cite{majumdar,danny1,danny2,carloBH,olaf}).

\item Our total area is taken to be given by $\sum_i j_i$ instead of the more usual loop gravity formula $\sum_i\sqrt{j_i(j_i+1)}$. This means that we are considering an equidistant area spectrum $j$ instead of the square-root of the $\SU(2)$ Casimir $\sqrt{j(j+1)}$. Actually these two spectrum only differ by ordering ambiguities \cite{krasnov} and are mathematically consistent. Our ansatz $\sum_i j_i$ is dictated by the fact that this the actual total area that is preserved by the $\U(N)$ transformation. Since it is natural to interpret this $\U(N)$ action as area-preserving diffeomorphisms, the natural choice is $\sum_i j_i$. This area spectrum has also been discussed in \cite{spain,hanno,lewandowski}.

\item A bigger issue is that $\dim_N[l]$ is counting many intertwiners with trivial legs $j_i=0$. In loop quantum gravity, this amounts to overcounting since spin network edges carrying a trivial $\SU(2)$ irrep are equivalent to no edge at all. We will solve this problem in the following section.

\item Another puzzle is that the standard black hole entropy formula only depends on the total area $l$. However, our number of intertwiners $\dim_N[l]$ depends on $l$ but also on the number of punctures $N$. We have two alternatives. The first possibility is to argue that the number of punctures $N$ is part of the bulk data: it depends on the spin network graph describing the quantum state of the exterior geometry. In this case, it would be natural that $N$ should be fixed in term of the area $l$ in the physical situation of a black hole (or isolated horizon). We discuss such a mechanism below. The second possibility is that we should simply sum over the number of punctures $N$ in order to remove this dependence. However, we have seen earlier that $\sum_N \dim_N[l]$ diverges. This is due to the over-counting of intertwiners with trivial legs, as we pointed out above. We will discuss in the next section how to remove these redundancies and how to implement a correct sum over $N$.

\end{itemize}

Despite these issues, we can give the asymptotics of the entropy $S_N[l]\,\equiv\ln\dim_N[l]$ for large area $l$ and number of legs $N$. We can easily compute it using Stirling approximation for the factorials. At leading order, the intertwiner dimension behaves as:
\be
\dim_N[l]\sim
\f{1}{2\pi lN^2}\left(1+\f{N}{l}\right)^{2l}\left(1+\f{l}{N}\right)^{2N-3}.
\ee
We distinguish three different asymptotic regimes:
\begin{itemize}

\item {\bf The Large Area Limit:}

We keep a fixed number of legs $N$, which we keep as an external parameter, and we send $l$ to infinity. In this case, we obtain a logarithmic entropy, which can not correspond to the black hole entropy:
\be
S_N[l]\underset{l\arr\infty}{\sim} (2N-4)\log l\,-(\ln (N-1)!+\ln (N-2)!) +\dots
\ee

\item {\bf The ``Continuum" Limit:}

We keep a fixed large area $l$ and we refine the outside graph in order to send $N$ to infinity. We obtain a linear growth of the entropy in term of $l$ but the proportionality factor diverges when $N$ grows large:
\be
S_N[l]\underset{N\arr\infty}{\sim}
2l\log N - \ln l! - \ln (l+1)! +\dots
\ee

\item {\bf The Linear Regime:}

The solution to match the area-entropy law for black holes is to assume that the number of punctures $N$ should depend on the horizon area $l$. This is a point of view similar to \cite{danny2} where it was argued that the combinatorics of the bulk spin network state should crucially depend on the considered physical context, i.e. that we are dealing with a black hole. Assuming that the number of punctures scales linearly with the area, $N\sim \lambda\,l$, it is straightforward to show that we recover a holographic behavior:
\be
S_N[l]\underset{N=\lambda l, \,l\arr\infty}{\sim}
2[(1+\lambda)\ln(1+\lambda)-\lambda\ln\lambda]\,l
-\,2\ln l\,
-\ln \,\f{2\pi(1+\lambda)}{\lambda}\,
+\dots
\ee
The leading order scales linearly with the area as expected and the proportionality coefficient depends on the parameter $\lambda$. Thus this construction provides an extra-parameter that we could fine-tune in order to recover the factor $\f14$ expected in the semi-classical regime, without having to fine-tune the Immirzi parameter (that we have omitted in our discussion since it does not enter the computation of $\dim_N[l]$). This proposal is similar to the bulk entropy proposal of \cite{danny2}, where there was an extra parameter (the bulk graph complexity) that could be fine-tune in the linear regime to match the exact area-entropy law without having to fix a precise value for the Immirzi parameter.

\end{itemize}

\subsection{The Binomial Transform and the Reduced Generating Functional}

We now would like to implement the sum over the number of punctures $N$. To this purpose, we need to remove the trivial legs of the intertwiners in order to avoid the over-counting responsible for the divergence of $\sum_N \dim_N[l]$.

The first step is to realize that the number of intertwiners between some spins $j_1,..,j_N$ only depends on the number of times each spin appears in the list. More precisely, considering the sequence of spins  $[j_1,..,j_N]$, we define the occurrence number $k_j$ for each spin $j\in\N/2$:
$$
[j_1,..,j_N]\,\arr\,\{(j,k_j)\} \qquad \textrm{with}\quad \left|\begin{array}{c}l=\sum_j jk_j \\ N=\sum_j k_j\end{array}\right.
$$
Then the key point is that the dimension of the intertwiner space $\dim_0[j_1,..,j_N]$ only depends on the occurrence number $k_j$ for $j>0$ and in particular does not depend on the number $k_0$ of times that the trivial irrep $j=0$ appears. This allows to decouple the trivial punctures from the counting and to define a number of intertwiners without trivial punctures:
\be
\dim_N[l]=\sum_{K=0}^N \bin{K}{N} D_K[l],
\qquad\textrm{with}\quad
D_K[l]=\sum_{\sum_{j\ge 1} k_j=K }
\f{K!}{\prod_j k_j!}\,\dim_0[\{k_j\}],
\ee
where $K=(N-k_0)$ is the number of non-trivial punctures. The new number $D_K[l]$ is the dimension of the space of intertwiners between $K$ non-trivial punctures with total area $l$. The binomial coefficients are statistical weight taking into account that punctures carrying the same spin are indistinguishable. Actually, $D_K[l]$ is mathematically called the {\it binomial transform} of $\dim_N[l]$ and we can inverse the previous relation~\footnotemark:
\be
D_K[l]=\sum_{N=0}^K (-1)^{K-N}\bin{N}{K}\dim_N[l].
\ee
\footnotetext{
Using this formula, we give explicitly these dimensions for low values of $N$:
$$
D_2[l]=1,\quad
D_3[l]=\f{(l-1)(l+4)}{2},\quad
D_4[l]=\f{(l-1)(l^3+9l^2+8l-36)}{12},
$$
$$
D_5[l]=\f{(l-2)(l-1)(l+3)(l+6)(l^2+9l-16)}{144}.
$$
}
Finally, we would like to define the total number of intertwiners with total area $l$ with arbitrary number of non-trivial punctures:
\be
D[l]\,\equiv\, \sum_{K\in\N} D_K[l] \,=\,\sum_{K=0}^{2l} D_K[l].
\ee
We notice that this sum is actually finite since the number of non-trivial punctures $K$ is automatically bounded by $2l$ since the minimal spin is $j=\f12$.

We introduce the generating functionals for the new dimensions $D_K[l]$:
\be
G_K(t)=\sum_l t^{2l}D_K[l],\qquad
\cG(u,t)=\sum_{K,l} u^Kt^{2l}D_K[l],\qquad
G(t)=\cG(1,t)=\sum_{l} t^{2l}D[l].
\ee
The relevance of the binomial transform is that the resulting generating functionals are easily related to the original ones:
\be
\cG(u,t)=\f{1}{1+u}\cF\left(\f{u}{1+u},t\right).
\ee
This can be seen from the integral representation  of these functionals. First, we have:
\beq
G_K(t)&\equiv&
\int dg\, \left[
\sum_j t^{2j}\chi_j(g)
\,-1\right]^K
\,=\,\sum_N(-1)^{K-N}\bin{N}{K}\,\int dg\, \left[\sum_j t^{2j}\chi_j(g)\right]^N\,,
\eeq
then we can compute the full generating functional by summing the previous expression over $K$ with the factor $u^K$:
\be
\cG(u,t)=\int dg\,\f{1}{1-u\left(\sum_j t^{2j}\chi_j(g) -1\right)}
\,=\,
\f2\pi\int_0^\pi d\theta\,\sin^2\theta\,\f{1-2t\cos\theta+t^2}{1-2t(1+u)\cos\theta+t^2(1+u)}.
\ee
Comparing this expression with the integral expression \Ref{cF} for $\cF(u,t)$ leads to the obvious relation between these two functionals. More directly, we could more simply use the identity on the binomial coefficients following from the series expansion of $(1-x)^{-k}$~:
\be
\f{x^K}{(1-x)^{K+1}}
\,=\,
\sum_{N\ge K} x^N \bin{K}{N},
\quad\textrm{in particular}\quad
\sum_{N\ge K} \f1{2^N} \bin{K}{N}=2.
\ee
Inserting this relation in the definition of the binomial transform leads to:
\be
\sum_N u^N\dim_N[l]
\,=\,
\sum_K D_K[l]\sum_{N\ge K}u^N\bin{K}{N}
\,=\,
\f{1}{1-u}\sum_K\left(\f{u}{1-u}\right)^KD_K[l],
\ee
which is exactly the formula that we wanted to prove. In particular, we have the relation between the number of intertwiners without trivial legs and the dimensions of the $\U(N)$ representation that we computed in the earlier section:
\be
\sum_N \f1{2^N}\dim_N[l]\,=\,2\sum_K D_K[l]\,=\,2D[l].
\ee
The statistical weight $1/2^N$ counterbalances the over-counting due to trivial legs of intertwiners in $\dim_N[l]$.

Now, using our previous calculations on $\cF$, we obtain an explicit expression for $\cG$ and n particular for $G(t)=\sum_{l} t^{2l}D[l]$:
\be
G(t)=\cG(1,t)=\f12\,\cF(\f12,t)=\f58+\f{1}{16t^2}\left(1-\sqrt{1-12t^2+4t^4}\right).
\ee
We can expand this functional around $t\sim 0$. Defining the variable $T\,\equiv t^2$, we get:
\be
\tG(T)\equiv G(t)= \sum_{l\in\N}D[l]T^{l}=1+T+6T^2+44T^3+360T^4+3152T^{5}+\dots
\ee
We can check that $\tG(T)$ satisfies the following differential equation:
\be
T(1-12T+4T^2)\pp_T\tG+(1-6T)\tG+(4T-1)=0.
\ee
This translates into a (very simple) recursion relation for the dimensions $D[l]$~:
\be
D[0]=D[1]=1,\qquad
D[l]=\f{1}{l+1}\,\left[6(2l-1)D[l-1]-4(l-2)D[l-2]\right],\quad\forall l\ge2.
\ee
Finally, either using the pole structure of $G(t)=\cG(1,t)$ or inserting the ansatz $D[l]\sim \alpha^l\,l^\sigma$ in the recursion relation, we obtain the asymptotic behavior of the intertwiner number $D[l]$, which is what we have been looking for.
Expanding the recursion relation to the next-to-leading-order correction (in $J^{\sigma-1}$), derive equations for $\alpha$ and $\sigma$~:
$$
\alpha^2-12\alpha+4=0,\qquad (2\sigma+3)=0.
$$
The value of $\alpha$ corresponds to the inverse of the convergence radius of the power series, as we have computed in the case of $\cF(u,t)$.

We finally obtain the asymptotic expression for this ``no-trivial puncture" entropy:
\be
S_{\o}[l]\equiv\log D[l]
\underset{l\arr\infty}{\sim}\,
l\ln\alpha\,-\f32 \ln l +\dots,\qquad
\alpha=\f1{\left(1-\sqrt{\f12}\right)^2}=6+4\sqrt{2}\simeq 11.6568.
\ee
This fits with the standard calculations (see e.g.\cite{spain}). We have a holographic leading order, and then we recover the usual loop gravity log-correction in $-\f32$ (for a more general discussion of the area-entropy law and its log-correction, see e.g.\cite{BHreview}). We also checked this expression numerically. Indeed numerical computations using the recursion relation are very fast. For $l=5000$, we get $S_{\o}[l]\simeq12\,265.1$, which agrees with our asymptotics with a $10^{-4}$ precision.

At the end of the day, we have shown how to consistently remove the trivial punctures and perform the sum over the number of punctures using the binomial transform tool. This has lead to a well-defined entropy satisfying the standard area-entropy law.

\section{From Intertwiners to Spin Networks}

Up to now, we have been discussing the $\U(N)$ structure of the intertwiner spaces. We would like to extend this $\U(N)$ point of view to spin network states.
To start with, let us remind the results obtained for intertwiners. Considering the intertwiner spaces
$$
\cH_{j_1,..,j_N}\,\equiv\, \textrm{Inv}[V^{j_1}\otimes..\otimes V^{j_N}],
$$
we have shown that the direct sum of such spaces for a fixed total area provides an irreducible representation of the unitary group $\U(N)$ with highest weight $[l,l,0,0,..]$:
$$
R^l_N \sim \bigoplus_{\sum_i j_i=l} \cH_{j_1,..,j_N}.
$$
It turns out that we can characterize these representations through group averaging. More precisely, we will show that the whole space of intertwiners with $N$ leg can be represented as a space of $L^2$ functions over a Grassmanian space $Gr_{2,N}$ defined as a quotient of $\U(N)$:
\be
\label{L2}
\cH_N
\,=\,
\bigoplus_{\{j_i\}}\cH_{j_1,..,j_N}
\,=\,
\bigoplus_l R^l_N
\,=\,
L^2\left(Gr_{2,N}\right),\quad \quad Gr_{2,N}\equiv \frac{\U(N)}{\U(N-2)\times \mathrm{SU}(2)}.
\ee
The subgroup $\U(N-2)\times \mathrm{SU}(2)$ stabilizes the highest weight vector $v$ with eigenvalues, $E_1\,v=E_2\, v\,=\,l\,v$ and $E_i\, v=0$ for all $i\ge 3$. The $\U(N-2)$ subgroup is generated by all the operators $E_{ij}$ with $i,j\ge 3$, while the SU$(2)$ subgroup is generated\footnotemark{} by the operators $(E_1-E_2), E_{12}, E_{21}$.

\footnotetext
{It is clear that $(E_1-E_2)v=0$ due to the choice of highest weight vector. Then we have $E_{12} v =0$ by definition of  the highest weight. Finally we can easily prove that $E_{21} v =0$ using the $\su(2)$ Lie algebra structure $[E_{12},E_{21}]=(E_1-E_2)$, or by the simple calculation:
$$
| E_{21} v |^{2} = v^{\dagger} E_{12}E_{21} v =
 v^{\dagger} [E_{12},E_{21}] v =  v^{\dagger} (E_{1}-E_{2}) v =0.
$$
}

To build the space of functions $L^2(Gr_{2,N})$, we first consider the Hilbert space $L^2(\U(N))$ and we impose a gauge invariance under $\U(N-2)\times \mathrm{SU}(2)$:
\be
\forall G\in\U(N),\, \forall H\in\U(N-2)\times \mathrm{SU}(2),\quad
f(GH)=f(G).
\ee
By the Peter-Weyl theroem, a basis of $L^2(\U(N))$ is provided by the matrix elements of all irreducible unitary representations of $\U(N)$. Such irreps are labeled by highest weight vectors $W=[l_1,..,l_N]$ where the $l_i$'s are arbitrary integers. Then a $L^2$ function has a unique decomposition:
\be
f(G)\,=\,
\sum_{W,a,b} f^{(W)}_{ab}\cD^{(W)}_{ab}(G),
\ee
where $a,b$ label a basis in the irrep with highest weight vector $W$, the $f^{(W)}_{ab}$ are the Fourier components of the function and finally the $\cD(G)$ are the matrix representing the group element $G$. Imposing the gauge invariance selects the irreps that have a vector invariant under $\U(N-2)\times \SU(2)$ and projects onto such vectors. This restricts the sum to irreps whose highest weight vector $W$ is of the type $[l,l,0,0,..]$~:
\be
f(GH)=f(G)
\,\Rightarrow\,
f(G)\,=\,
\sum_{l,a} f^{(W_{l})}_{a}\cD^{(W_{l})}_{aW_{l}}(G),\quad W_{l}=[l,l,0,0,..],
\ee
where the highest weight is of the expected form, the vector $b$ has been projected on the highest weight vector $W$ and $a$ still labels a basis of each irrep.  This provides the isomorphism \Ref{L2} between the space of $\SU(2)$ intertwiners with $N$ legs and the $L^2$ space of gauge invariant functions on $\U(N)$. For instance, the space of 4-valent intertwiners is isomorphic to:
\be
\cH_{N=4}=L^2\left(\frac{\U(4)}{\U(2)\times \mathrm{SU}(2)}\right).
\ee
This space is $9$-dimensional. It is related to the space of classical 3d tetrahedron with fixed total area, which are labeled by 5 numbers (the edge lengths for example up to a scale) plus 4 phases associated to each face.

More generally the space $Gr_{2,N}$ is associated to the space of polyhedra with $N$ faces and fixed total area. More precisely we define $P_{N}$ to be the space of convex polyhedra possessing $N$ faces,
$3(N-2)$ edges and $2(N-2)$ vertices.
This can be equivalently described as the space of polyhedra with $N$ faces and such that the
vertices are all trivalent. Since the relation $3V=2E$ and the spherical condition  $ F-E+V =2$ determines $E$ and $V$.
The dimension of this space is equal to the number of edges hence is equal to $3(N-2)$.
If one go from $N$ to $N+1$ one sees that the number of edges increase by $3$ while the number of
vertex increase by $2$. This is exactly what happens when we perform a $1$-to-$3$ move on the boundary, this is a move that blows up a vertex into a triangle. Such a move can be realized by cutting a polyhedra in $P_{N}$ with a new plane around a vertex and taking the convex envelop.
Differently shaped polyhedra in $P_{N}$ can be obtained from one another by action of a succession  of $2$-to-$2$ exchange moves that do not change the number of edges faces and vertices.
If we denote the space of Polyhedra in $P_{N}$ having a fixed total area $A$ by $P_{N}(A)$.
We expect the isomorphism
\be
P_{N}(A) = \U(1)^{N}\backslash Gr_{2,N}\, \quad \mathrm{or} \quad Gr_{2,N} = P_{N}(A) \times U(1)^{N}.
\ee
Both spaces ($Gr_{2,N}$ and  $P_{N}(A)\times U(1)^{N}$) have dimension $4N- 7 = 3(N-2) -1 +N $.
We postpone a more detail study of the spaces $Gr_{2,N}$ and their geometrical meaning for future investigation.

Now that we have described the Hilbert space of intertwiners, we can glue them together in order to describe the space of spin network functionals. Considering a particular (oriented) graph $\Gamma$, we define the Hilbert space of spin networks as $L^2$ functions of one group element $g_e\in\SU(2)$ per edge $e\in\Gamma$ that are invariant under the $\SU(2)$ action at each vertex:
\be
\vphi\in L^2(\SU(2)^E/\SU(2)^V),\quad
\forall h_v\in\SU(2)^V,\,\vphi(\{g_e\})=\vphi(\{h_{s(e)}^{-1}g_eh_{t(e)}\}),
\ee
where $E$ is the number of edges, $V$ the number of vertices, and $s(e),t(e)$ respectively the source and target vertices of the (oriented) edge $e$. A basis of this space is provided by applying the Peter-Weyl theorem to $\SU(2)^E$. We label all edges by a $\SU(2)$ irreducible representation $j_e\in\N/2$ and we attach an intertwiner (basis) state to each vertex:
\be
\cH_\Gamma=L^2(\SU(2)^E/\SU(2)^V)
\,=\,
\bigoplus_{\{j_e\}}\bigotimes_{v\in\Gamma}\cH_{j_1^v,..,j_{N_v}^v},
\ee
where $N_v$ counts the number of edges at the vertex $v$ and $j_1^v,..,j_{N_v}^v$ are the spins attached to these edges.
Having started with the degrees of freedom $g_e$ attached to the edges, we have shifted the perspective to attaching the degrees of freedom to the intertwiners on the vertices. We can conclude this shift of viewpoint using our description of the intertwiner spaces. For each vertex, we have:
$$
\cH_{N_v}=L^2(Gr_{2,N_v}),\quad Gr_{2,N_v}\,=\,\U(N_v)/\U(N_v-2)\times \mathrm{SU}(2).
$$
To glue these spaces together, we simply have to impose that the spin $j_e$ is the same for the two vertices $s(e)$ and $t(e)$. This is done by imposing the two operators $E_e^{s(e)}$ and $E_e^{t(e)}$, measuring the value of the spin $j_e$ in the space of intertwiners respectively attached to $s(e)$ and $t(e)$, have the same value. This amounts to an extra $U(1)$ invariance for each edge:
\be
\cH_\Gamma=L^2(\SU(2)^E/\SU(2)^V)
\,=\,
L^2\left((\times_e\U_{(e)}(1))\backslash(\times_v Gr_{2,N_v})\right),
\ee
where $\U_{(e)}(1)$  is generated by $E_e^{s(e)}-E_e^{t(e)}$, and the quotient is take on the left hand side, i.e we require the functions to satisfy:
\be
\left|\begin{array}{ll}
\forall H_v\in\U(N_v-2)\times  \mathrm{SU}(2),\,&
f(\{K_v\})\,=\,
f(\{K_v H_v\}) \\
\forall T_e\in\U_{(e)}(1),\,&
f(\{K_{s(e)},K_{t(e)},K_v\})\,=\,
f(\{T_eK_{s(e)},T_eK_{t(e)},K_v\}).
\end{array}
\right.
\ee
This establishes an isomorphism between the space of spin network functionals based on a graph, which are functions of one $\SU(2)$ group element per edge, and a space of functions of one $\U(N)$ group element per vertex. We believe that such a edge-vertex duality should be useful to study dynamical aspects of loop quantum gravity, when looking at deformations of the geometry defined by the spin network states.

\section{Generalizations}

We have studied the space of $\SU(2)$ intertwiners with $N$ legs and identified the action of $\U(N)$ on that space. This procedure can actually be generalized beyond the context of a $\SU(2)$ gauge theory. We propose three possible extensions:
\begin{itemize}


\item {\bf Enlarging the $\SU(2)$ Gauge Group to $\SU(d)$~:}

As we have seen earlier, the Schwinger representation of $\SU(2)$ in term of harmonic oscillators works for any unitary group $\U(d)$ (and $\SU(d)$). This requires $d$ uncoupled oscillators. Then studying intertwiners with $N$ legs, we work with a double series of $N\times d$ oscillators which naturally carries a representation of $\U(N)$. As we have shown in section \ref{proofweight}, this leads to highest weight representations of $\U(N)$ with $d$ non-trivial eigenvalues when $d<N$ and to arbitrary representations of $\U(N)$ as soon as $d\ge N$. Working out the details of the intertwiner requirement will certainly lead to constraints on the highest weight. This generalization could be applied to gauge theories. It shows that we'll always have this $\U(N)$ action which can be interpreted as ``area"-preserving diffeomorphisms.
This procedure should actually work for arbitrary Lie group whose algebra can be formed from harmonic oscillators.


\item {\bf Going Super-Symmetric~:}

We can also try to apply our method to super-symmetric theories. For instance $\osp(1|2)$ can be realized by adding one fermionic oscillator $c$ to the Schwinger representation of $\SU(2)$ \cite{OH}:
$$
[a,a^\dag]=[b,b^\dag]=\{c,c^\dag\}=1,
$$
$$
J_z=\f12(a^\dag a-b^\dag b),\,J_+=a^\dag b,\quad
Q_+=\f12(a^\dag c+c^\dag b),\,Q_-=\f12(a c^\dag-c b^\dag).
$$
Then considering intertwiners with $N$ legs, we take $N$ copies of this algebra, with the fermionic oscillators anti-commuting with each other $\{c_i,c_j\}=0$.. Once again, we can build quadratic invariant operators, $E_{ij}=a_i^\dag a_j+b_i^\dag b_j+c_i^\dag c_j$. These $E_{ij}$ still form a (bosonic) $\U(N)$ algebra which commutes with the global $\osp(1|2)$ action. It could be interesting to see what $\U(N)$ representations it leads to and compare it with our results for $\SU(2)$. This would explain how the presence of a supersymmetric fermion deforms the action of the area-preserving diffeomorphisms.


\item {\bf Quantum Deformation and Cosmological Constant~:}

Finally, we should consider the quantum deformation of $\SU(2)$, or of any arbitrary unitary group $\U(d)$. In the loop quantum gravity context, this usually corresponds to the presence of a non-vanishing cosmological constant. The harmonic oscillator construction still works for $q$-deformation of $\U(d)$ using $q$-oscillators. More precisely, considering a double series of $N\times d$ of $q$-oscillators, we can build representations of $\U_q(d)\times \U_q(N)$ (at the level of the Lie algebra and for the $R$-matrix) \cite{quesne}.

\end{itemize}

\section{Conclusions and Outlook}


We have studied the Hilbert space of $\SU(2)$ intertwiners with $N$ legs of loop quantum gravity from the viewpoint of the $\U(N)$ structure encovered in \cite{OH}. We actually showed that the space of $\SU(2)$ intertwiners with $N$ legs with fixed total area defined as the sum of spins $\sum_i j_i$ provides an irreducible representation of $\U(N)$. We have moreover identified the  highest weights of these representations and showed it correspond to bivalent intertwiners. This work allowed us to interpret this $\U(N)$ action as the area-preserving diffeomorphisms acting on the (topologically spherical) boundary surface dual to the intertwiner. We have further explained how this structure generalizes to the space of spin network states based on some fixed graph. Thus these discrete area-preserving diffeomorphisms will certainly be relevant to understanding how to deform the quantum geometry of spin network states in loop quantum gravity and how space-time diffeomorphisms should arise in the continuum/semi-classical regime of the theory.

Identifying the space of $\SU(2)$ intertwiners with $N$ legs with fixed total area as a $\U(N)$ representation also allowed us to compute in a simple way the dimension of this space using the standard hook formula of the representation theory of $\U(N)$. We provided an alternative computation to check that this result was indeed right, confirming that our framework is mathematically consistent. A side-product is an actual calculation of the black hole entropy. Up to a subtlety on (over)counting trivial legs of the intertwiners, we computed the corresponding generating functional and recovered the standard asymptotics for the black hole entropy in the large area regime as one would obtain from counting quantum states in the $\SU(2)$ Chern-Simons theory.

The present framework is a stem which we could develop at least in two directions. On the one hand, we should study further the geometrical interpretation of this $\U(N)$ action which has been sketch in the last section. One way would be to look at this $\U(N)$ action on semi-classical intertwiner states which should describe semi-classical geometries on the boundary surface (see appendix \ref{acoh1} and \ref{acoh2} fore more details). Such semi-classical states have already been investigated and defined as holomorphic intertwiners \cite{Holomorph}. They correspond to three-dimensional polyhedra and we could see how the $\U(N)$ transformations deforms these polyhedra. On the other hand, this $\U(N)$ structure can be understood in term of matrices and the intertwiner dynamics (and the corresponding surface dynamics or black hole dynamics) be interpreted in term of matrix models. Since matrix models have an underlying conformal symmetry, this would open the door to another link between the loop quantum gravity dynamics and conformal field theory. A last speculation is that such a relation might also allow to identify an integrable sector of the LQG dynamics.

\section*{Acknowledgments}

EL is partially supported by the ANR ``Programme Blanc" grants LQG-06 and LQG-09.
Research at Perimeter Institute is supported by the Government of Canada through Industry Canada and by the Province of Ontario through the Ministry of Research and Innovation.

\appendix

\section{$\U(N)$ Characters and Generating Functionals}

We can go further and generalize the intertwiner counting by considering the characters of $\SU(N)$.
The characters give the trace of unitary transformations in the considered representation and we compute it on the (abelian) Cartan subgroup of $\U(N)$, i.e diagonal group elements. Indeed, any group element can be diagonalized and is conjugated to an element of the type $\exp(i(s_1E_1+..+s_NE_N))$. The characters in the irreducible representation~\footnotemark{ } with highest weight $[l_1,l_2,..,l_N]$, $l_1\ge l_2\ge..\ge l_N$, is given as a quotient of Van der Monde determinants in term of the variables $t_i=\exp(is_i)$:
\be
\chi_{[l_i]}(t_1,..,t_N)\,=\,
\f{\det (t_k^{l_i+N-i})_{ik}}{\det (t_k^{N-i})_{ik}}.
\ee
\footnotetext{
If we define the highest weight $[l_1,l_2,..,l_N]$ with the opposite convention $l_1\le l_2\le..\le l_N$, the expression of the character is slightly different:
$$
\chi_{[l_i]}(t_1,..,t_N)\,=\,
\f{\det t_k^{l_i+i-1}}{\det t_k^{i-1}}.
$$
}
We now focus on our case where the highest weight is $[l,l,0,0,..]$ and we label the characters with simply $l$. First, one can easily check that this formula leads back to the formula for the dimension $\dim_l[N]$ where all $s_i$ are sent to 0 or equivalently all $t_i$ are sent to 1.
Then the $\U(N)$ character is related to the dimensions of the $\SU(2)$ intertwiner spaces:
\be
\chi_l(t_1,..,t_N)\,=\,
\sum_{j_1+..+j_N =l}t_1^{2j_1}..t_N^{2j_N}
\dim_0[j_1,..,j_N].
\ee
We can further sum over the representation label $l$ and define a general generating functional:
\beq
F_N(t_1,..,t_N)&=&
\sum_J\chi_J(t_1,..,t_N)
=\sum_{\{j_i\}}\prod_it_i^{2j_i}\,
\dim_0[j_1,..,j_N] \\
&=&\f2\pi\int_{-1}^{+1} dx\,\f{\sqrt{1-x^2}}{\prod_i^N(1-2t_ix+t_i^2)}.\nn
\eeq
This reduces to the previous generating functional when all $t_i$ are taken equal, $F_N(t)=F_N(t,..,t)$. These integrals can be computed as before. For instance, for 4-valent intertwiners, $N=4$, we obtain:
\be
F_4(t_1,..,t_4)=\f{1-\prod_i t_i}{\prod_{i<j}(1-t_it_j)}.
\ee
This simple expression does not generalize straightforwardly to higher values of $N$.

\section{From Harmonic Oscillators to $\SU(2)$ Coherent States}
\label{acoh1}

Using the Schwinger representation, we can build the coherent states for $\SU(2)$ from coherent states for the system of two uncoupled oscillators:
$$
|z_a,z_b\ra\,=\,
e^{-\f12(|z_a|^2+|z_b|^2)}\,
\sum_{n_a,n_b}\f{z_a^{n_a}z_b^{n_b}}{\sqrt{n_a!n_b!}}\,|n_a,n_b\ra_{OH}.
$$
These coherent states are normalized, $\la z_a,z_b|z_a,z_b\ra=1$, and the expectation values of the $J$-operators are easy to compute:
\be
\la \cE\ra=\f12(|z_a|^2+|z_b|^2),\quad
\la J_z\ra=\f12(|z_a|^2-|z_b|^2),\quad
\la J_+\ra=\bar{z}_az_b,\quad
\la J_-\ra=z_a\bar{z}_b.
\ee
Defining the complex ratio $z\,\equiv z_b/z_a$, we can re-write these mean values as:
\be
\la J_z\ra=\la \cE\ra\,\f{1-|z|^2}{1+|z|^2},\quad
\la J_+\ra=2\la \cE\ra\,\f{z}{1+|z|^2},\quad
\la J_-\ra=2\la \cE\ra\,\f{\bar{z}}{1+|z|^2},
\ee
for which it is easy to check that $\la \vec{J}\ra\cdot\la \vec{J}\ra=\la \cE\ra^2$. This is almost the same as for the $\SU(2)$ coherent states. To get the exact $\SU(2)$ coherent states, we simply need to project the oscillators' coherent states
on the space with fixed total energy. Indeed, fixing the eigenvalue of the $\cE$-operator correspond to fixing the spin $j$ amounts to restricting the sum defining the coherent states to energy levels satisfying $n_a+n_b=2j$, and we get:
\be
P_j|z_a,z_b\ra\,=\,
e^{-\f12(|z_a|^2+|z_b|^2)}\,z_a^{2j}|j,z\ra,\qquad
\textrm{with}\quad
|j,z\ra=\sum_{m=-j}^{+j}\f{z^{j-m}}{\sqrt{(j-m)!(j+m)!}}\,|j,m\ra.
\ee
Up to the normalization of $|j,z\ra$, we recognize the $\SU(2)$ coherent states with the usual expectation values for the $J$-operators:
\be
\la j,z|j,z\ra= \f{(1+|z|^2)^{2j}}{(2j)!},\quad
\la J_z\ra=j\,\f{1-|z|^2}{1+|z|^2},\quad
\la J_+\ra=\f{2jz}{1+|z|^2},\quad
\la J_-\ra=\f{2j\bar{z}}{1+|z|^2}.
\ee
The projector $P_j$ fixing the spin $j$ can be written as a complex contour integral along the unit circle:
\be
P_j|z_a,z_b\ra=
\f1{2i\pi}\oint_{\cS_1}\f{d\lambda}{\lambda}\,
\lambda^{-2j}\,|\lambda z_a,\lambda z_b\ra,
\ee
\be
e^{-\f{|z_a|^2}2(1+|z|^2)}\,|j,z\ra\,=\,
\f{1}{2i\pi}\oint_{\cS_1}\f{d\lambda}{\lambda}\,
(\lambda z_a)^{-2j}\,|\lambda z_a,\lambda zz_a\ra,
\quad\forall z_a.
\ee
Note that the multiplication by $\lambda \in\cS_1$ is achieved by a simple action of $\cE$: indeed, we have as usual $e^{2i\alpha \cE}\,|z_a,z_b\ra=|e^{i\alpha}z_a,e^{i\alpha}z_b\ra$.

\medskip

We can also give the formula for the resolution of the identity on the two-oscillator Hilbert space in term of the ``projected" coherent states $|j,z\ra$. Indeed, starting with the standard formula,
$$
\mathbbm{1}_{\cH_a\otimes\cH_b}=
\sum_{n_a,n_b}|n_a,n_b\ra\la n_a,n_b|=
\f1{\pi^2}\int d^2z_ad^2z_b\,|z_a,z_b\ra\la z_a,z_b|.
$$
Since the full Hilbert space $\cH_a\otimes\cH_b$ decomposes as the direct sum of the Hilbert spaces at fixed energy $n_a+n_b=2j$ , we can insert the projections operators $P_j$ in this resolution of the identity. Then performing the change of variable from $d^2z_ad^2z_b$ to $|z_a|^2d^2z_ad^2z$, we obtain:
$$
\mathbbm{1}=
\f1{\pi^2}\sum_j\int d^2z_ad^2z_b\,P_j|z_a,z_b\ra\la z_a,z_b|P_j =
\f1{\pi^2}\sum_j\int d^2z d^2z_a\,|z_a|^{2(2j+1)}e^{-|z_a|^2(1+|z|^2)}\,|j,z\ra\la j,z|.
$$
Finally, we can compute the integral over $z_a$ and introduce normalized coherent state $|j,z\ra_n=\f{\sqrt{(2j)!}}{ (1+|z|^{2})^{j}}|j,z\ra $. This gives a very simple final formula:
\be
\mathbbm{1}=
\f1{\pi}\sum_j\int d^2z\, \f{(2j+1)!}{(1+|z|^2)^{2j+2}} \,|j,z\ra\la j,z|=
\f1{\pi}\sum_j (2j+1)\int \f{d^2z}{(1+|z|^2)^{2}} \,|j,z\ra_n{}_n\la j,z|,
\ee
where we recognize the resolution of the identity for $\SU(2)$ coherent states with the invariant measure $d^2z/{(1+|z|^2)^{2}}$ on the 2-sphere.

\section{From $\SU(2)$ Coherent States to Coherent Intertwiners}
\label{acoh2}

To write coherent intertwiner states that would correspond to semi-classical chunks of volume with a well-defined surface boundary, the simplest strategy is to tensor together coherent states for $\SU(2)$ and to group average over the global $\SU(2)$ action to project on the intertwiner space \cite{Coh1,Coh2}. Thus considering an intertwiner with $N$ legs, we tensor together $N$ coherent states:
\be
|j_1,z_1,..,j_N,z_N\ra_0\,\equiv\,
\int_{\SU(2)}dg\,
g\rhd\left(\otimes_i |j_i,z_i\ra\right)
\,=\,
\int_{\SU(2)}dg\,
\otimes_i g|j_i,z_i\ra.
\ee
The action of the $E_{ij}$-operators commute with this action of $g$ since they are invariant operators by definition:
\be
E_{ij}|j_1,z_1,..,j_N,z_N\ra_0
\,=\,
\int_{\SU(2)}dg\,
g\rhd\left(E_{ij}\otimes_k |j_k,z_k\ra\right).
\ee
The big difference between the scalar product operator $J_{(i)}\cdot J_{(j)}$ and the quadratic invariant operators $E_{ij}$ is that the scalar product operators do not change the $\su(2)$ representations and act on the intertwiner space with fixed labels $j_1,..,j_N$ while the $E_{ij}$'s induce shifts in $j_i$ and $j_j$. Indeed the annihilation operators $a_i$ and $b_i$ will lower $j_i$ while the creation operators $a_j^\dag$ and $b_j^\dag$ will increase $j_j$. This is the price to pay in order to have quadratic operators instead of quartic.


The next step is to re-parameterize coherent intertwiners as holomorphic intertwiners in term of cross-ratios $Z_1,..,Z_{N-3}$ which allows to factor out the global $\SU(2)$ invariance. In \cite{Holomorph}, it was shown how this space of holomorphic intertwiners corresponds to classical tetrahedra for $N=4$. Then one could investigate how the action of  $\U(N=4)$ actually deforms classical tetrahedra. This would help understanding the precise geometrical interpretation of $\U(N)$ as area-preserving diffeomorphisms. Of course, for arbitrary values of $N$, one should first repeat the analysis of \cite{Holomorph} and show how holomorphic intertwiners can be mapped generically to classical polyhedra.



\end{document}